\documentclass[pre,preprint,aps,floatfix,showpacs,showkeys]{revtex4}

\usepackage{graphicx}

\begin{document}

\title{Cross-country hierarchical structure and currency 
crisis}
\thanks{In press in Internationaal Journal of Modern Physics C}

\author{Guillermo J. Ortega}
\affiliation{Department of Physics, Universidad de Buenos Aires and CONICET
Ciudad Universitaria, Pabell\'on I, 1428, Buenos Aires, Argentine\\
ortega@df.uba.ar}

\author{David Matesanz}
\affiliation{Department of Applied Economics, Universidad de Oviedo
Avda. Cristo s/n, 33006, Oviedo, Spain\\
matesanzdavid@uniovi.es}

\begin{abstract}
Using data from a sample of 28 representatives countries, we 
propose a classification of currency crises consequences based 
on the ultrametric analysis of the real exchange rate movements 
time series, without any further assumption. By using the matrix 
of synchronous linear correlation coefficients and the appropriate 
metric distance between pairs of countries, we were able to construct 
a hierarchical tree of countries. This economic taxonomy provides 
relevant information regarding liaisons between countries and a 
meaningful insight about the contagion phenomenon.

\end{abstract}

%\thanks{In press in Internationaall Journal of Physics C}

\maketitle

\section{Introduction}

Prediction and contagion of financial crises have 
received much attention in recent years. The financial 
instability during the nineties has caused intense exchange 
and banking crises, in developed and, especially, in developing 
countries. Most of the empirical literature has focused their interest 
in identification, prediction and contagion of currency crises, and 
the macroeconomic variable which seems to better account for both 
effects is the real exchange rate (RER) \cite{KL98,Ab03,Pe05,PA03}. 
However, conclusive results 
from the empirical literature are hard to achieve. 
One of the reason for unconvincing answers in this debate is
the enormous differences in periods and 
countries used in the empirical works, without taking account for regional 
or country specific differences in the underlying dynamics of the 
variables time series used \cite{Zh01} \cite{PA03}. 

From a methodological point of view, techniques and tools 
formerly used in the physical and biological fields, have become 
to be applied in the analysis of economic data \cite{MS00,BP00}, 
in particular, 
to the case of stock portfolios. In this case, correlation based 
clustering of synchronous financial data has been performed to obtain a 
taxonomy of a set stocks from the US equity market. The last objective 
in this kind of works is to improve economic forecasting and modeling the 
complex dynamic underlying the raw data and their basic hypothesis is that 
financial time series are carrying valuable economic information 
that can be detected.

Following the above ideas, we shall extract information present in the 
correlation matrix of the RER in a sample of 28 representative countries, 
in the period of 1990-2002. By using the subdominant ultrametric 
associate with a metric distance in the correlation space, we first 
construct the Minimum Spanning Tree (MST) which provides a topological 
picture of the countries links. Then, we shall proceed to construct a 
hierarchical tree associated with the distance matrix in order to 
obtain a country taxonomic description provided by the real exchange data. 
So, the main aim of this work is to detect hierarchical structure of our 
country sample that arises from the relations links in their exchange 
rate dynamics. Clustering countries in such a way could be of importance 
in several economic aspects related to the empirical currency crises and 
contagion literature. Probably the most important is the identification 
of homogenous countries in their exchange rate dynamics in order to 
construct better regional Early Warning Systems (EWS), more accurate 
forms of dating the events 
of crises especially design for homogeneous regions (or isolated countries) 
and, finally, for understanding the possibilities of forecasting of contagion. 

\section{Methodology}

\subsection{Data}
\begin{figure}
\includegraphics[width=10cm,angle=270]{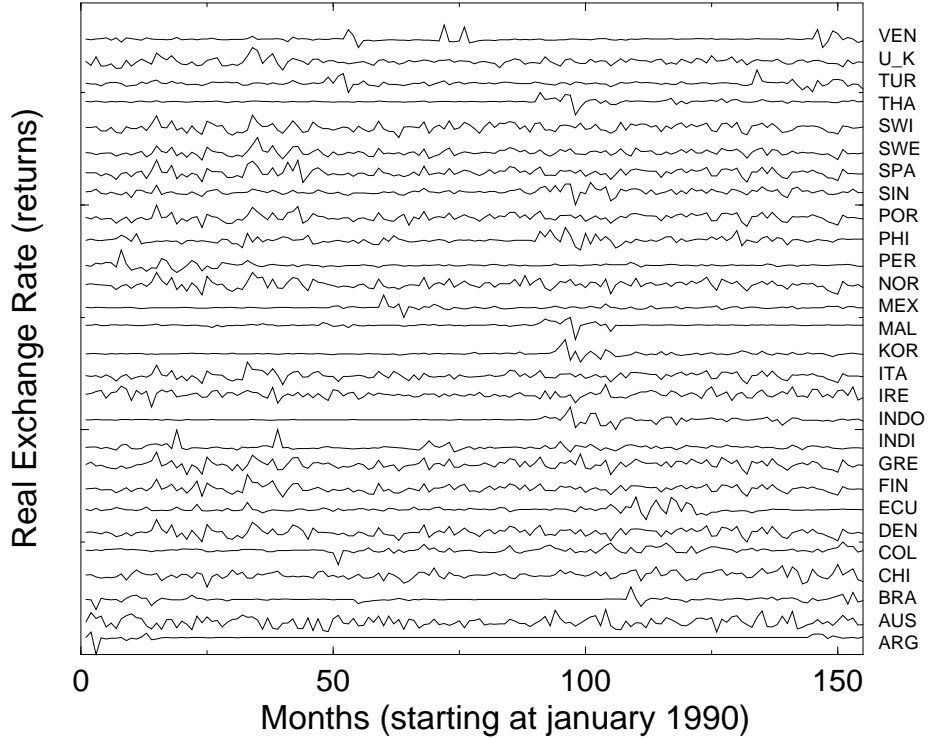}
\caption{
Returns of real exchange rate time series in the 28 countries.
Countries are ordered alphabetically from bottom to top. Monthly data from
January 1990 to December 2002 has been used. Countries
are labeled accordingly with the symbols listed in the Appendix.
\label{f1}}
\end{figure}

Returns from $RER$ in each of the 28 time series has been 
calculated in the usual way, 
\begin{equation}
rRER_i(k) = \frac{RER_i(k+1) - RER_i(k)} {RER_i(k)}
\label{eq1}
\end{equation}
where $RER_i(k)$ is the monthly real exchange rate from 
country $i$, at month $k$, and $rRER_i(k)$ the corresponding
return.
The period 1990-2002 has been used, yielding a total of 156 data
points for each country. Figure (\ref{f1}) shows the actual time series
used for further calculations. $RER$ is computed as the ratio of foreign
price proxied by U.S. consumer price to domestic consumer price, and the
result is multiplied by the nominal exchange rate of the domestic currency with
U.S. dollar. Data has been drawn
from International Financial Statistics in the IMF database available
on-line $(http://ifs.apdi.net/imf/logon.aspx)$.

\subsection{Numerical Methods}
In order to quantify the degree of similarity between pairs
of $RER$ time series belonging to different countries, we have calculated
the Pearson correlation coefficient \cite{PT92}
\begin{equation}
\rho_{i,j} = \frac{\sum_{k=1}^{N} (rRER_{i}(k)-<{rRER_i}>)
(rRER_{j}(k)-<{rRER_j}>)}
{{\sqrt{\sum_{k=1}^{N} {(rRER_{i}(k)-<{rRER_i}>)}^2}}
{\sqrt{\sum_{k=1}^{N} {(rRER_{j}(k)-<{rRER_j}>)}^2}}}
\label{eq2}
\end{equation}
where $<rRER_i>$ is the mean value of $rRER_i$ in the period
considered. Because $\rho_{i,j}$ is a measure of similarity,
and a measure of ''distance" is actually needed in order to 
construct the ultrametirc space \cite{RT86}, following
Gower \cite{Go66}, we define the distance between the
time evolution of $rRER_i$ and $rRER_j$ as,
\begin{equation}
d(i,j) = \sqrt{ \rho_{i,i} + \rho_{j,j} - 2 \rho_{i,j}} =
\sqrt{2 ( 1 -  \rho_{i,j})}
\label{eq3}
\end{equation}
The last equality came from the symetry property of the
correlation matrix, $\rho_{i,j} = \rho_{j,i}$ and the 
normalization $\rho_{i,i} = 1$  $\forall i$. In
this way, $d_{i,j}$ 
fulfils the three axioms of a distance: 
\begin{itemize}
 \item $d(i,j) = 0$ if and only if $i=j$
 \item $d(i,j) = d(j,i)$
 \item $d(i,j) \leq d(i,l) + d(l,j)$
\end{itemize}
The third axiom, the triangular inequality, characterize a metric
space. An {\it ultrametric space}, on the other hand, is endowed with
a distance that obeys a stronger inequality, the
{\it ultrametric distance} $d(i,j)^{<}$:\\
\begin{equation}
d(i,j)^{<} \leq max\{ d(i,l),d(l,j) \}
\label{eq4}
\end{equation}
Thus, it follows that the distance matrix given by Equation (\ref{eq3})
satisfies ultrametricity and a hierarchical tree can be uniquely
constructed \cite{RT86}. 

One method to obtain $d(i,j)^{<}$ directly from the distance
matrix $d(i,j)$ is through the MST method \cite{RT86}.
Given the metric space $(\Omega, d)$, that is, countries 
and the distance defined by Equation (\ref{eq3}), there is
associated with this space a nondirected graph with the same 
elements of $\Omega$
as vertices, and links between the elements $(i,j)$, the
distances $d(i,j)$. The MST is a tree with the same vertices 
as in $\Omega$ but of minimal total lenght. Although more than
one MST can be constructed on $\Omega$, $d^{<}$ is unique.
With the information provided by the MST, the distance 
$d(i,j)^{<}$ between two elements $i$ and $j$
in $\Omega$ is given by
\begin{equation}
d(i,j)^{<} = max\{d(w_i,w_{i+1}), 1 \leq i \leq n-1\}
\label{eq5}
\end{equation}
where $C_{i,j} = \{(w_1,w_2),(w_2,w_3),...,(w_{n-1},w_{n})\}$
denotes the unique path in the MST between $i$ and $j$ 
($w_1 = i, w_n = j$). We shall show in the next section how
to construct $d(i,j)^{<}$ in our particular case.

In what follows, we shall follow closely the analysis and methodology
done in the work of R. Mantegna \cite{Ma99} in the case of stocks.
A comprehensive review of ultrametricity, hierarchical trees and
clustering methods can be found in reference \cite{RT86}

\section{Results}
\begin{figure}
\includegraphics[width=10cm,angle=270]{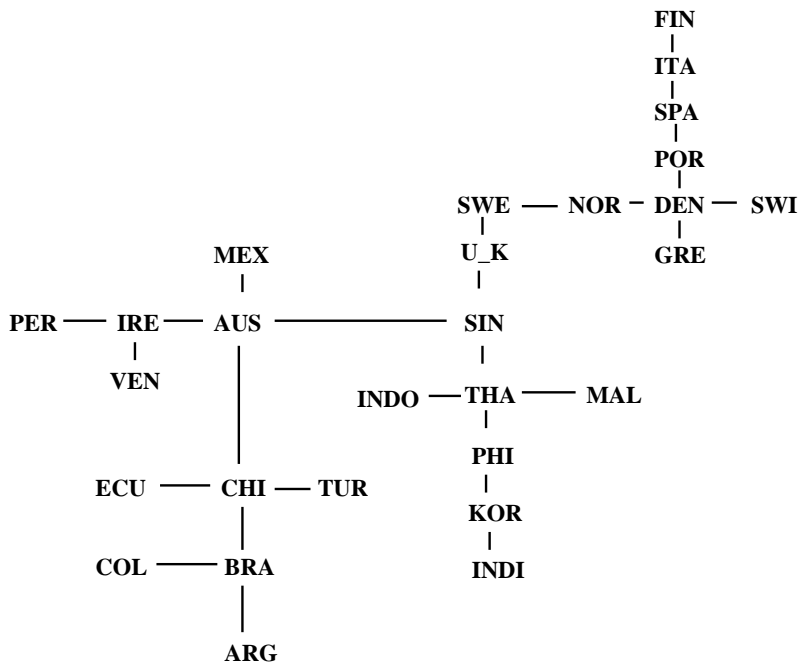}
\caption{
Minimal Spanning Tree connecting the 28 countries. Countries
are labeled accordingly with the symbols listed in the Appendix.
\label{f2}}
\end{figure}

We first construct the MST directly from the
distance matrix $d(i,j)$.
One begins by connecting
the closest countries given by $d(i,j)$, in this case POR-SPA with
a distance equal to 0.41. Table \ref{tab1} shows some representative
distances. One then proceeds by linking the remaining
countries accordingly with their closeness to the previously
connected countries. For instance, in the distance matrix, the shortest distance
following POR-SPA is DEN-SWI with a distance of 0.411, and in this
way, we have another link between both countries. The next one is
DEN-GRE with a distance of 0.464. We then proceed to connect GRE to
the former pair DEN-SWI, giving GRE-DEN-SWI. At this moment, we
have two "clusters", POR-SPA and GRE-DEN-SWI. Proceeding in the
above explained way, we finally construct a tree with the 28 countries
and 27 links among them. Figure (\ref{f2}) shows the complete
MST given by the distance matrix $d(i,j)$.

%\begin{table}[ht]
\begin{table}[t]
\caption{Some representatives distances 
between pairs of countries, {\it i.e.} $d(i,j)$}
{\begin{tabular}{|l|c|c|} \hline
Distance & country & country \\
\hline
0.410 &  POR\* &  SPA\* \\ \hline
0.411 &  DEN\* &  SWI\* \\ \hline
0.464 &  DEN   &  GRE\* \\ \hline
0.465 &  DEN   &  NOR\* \\ \hline
0.490 &  DEN   &  POR  \\ \hline
 ...  &  ...   &  ...  \\ \hline
0.666 &  MAL\* &  THA\* \\ \hline
0.669 &  ITA   &  NOR  \\ \hline
0.669 &  FIN   &  SWI  \\ \hline
 ...  &  ...   &  ...  \\ \hline
0.797 &  SIN\* &  THA  \\ \hline
0.834 & INDO\* &  THA  \\ \hline
0.847 &  MAL   &  SIN  \\ \hline
0.905 &  PHI\* &  THA  \\ \hline
0.926 &  INDO  &  SIN  \\ \hline
0.937 &  INDO  &  MAL  \\ \hline
0.952 &  SWE   &  U\_K\* \\ \hline
0.972 &  ITA   &  U\_K  \\ \hline
 ...  &  ...   &  ...  \\ \hline
1.020 &  AUS\* &  IRE  \\ \hline
 ...  &  ...   &  ...  \\ \hline
1.137 &  ARG\* &  BRA\* \\ \hline
1.156 &  PHI   &  SIN  \\ \hline
1.171 &  BRA   &  CHI\* \\ \hline
1.184 &  KOR\* &  PHI  \\ \hline
 ...  &  ...   &  ...  \\ \hline
1.241 &  BRA   &  COL\* \\ \hline
 ...  &  ...   &  ...  \\ \hline
1.272 &  CHI   &  TUR\* \\ \hline
 ...  &  ...   &  ...  \\ \hline
1.281 &  AUS\* &  MEX\* \\ \hline
 ...  &  ...   &  ...  \\ \hline
1.329 & INDI\* &  KOR  \\ \hline
 ...  &  ...   &  ...  \\ \hline
1.330 &  SPA   &  TUR  \\ \hline
\end{tabular}
\label{tab1}}
\end{table}

Armed with the information provided by the distance matrix $d(i,j)$
and the MST,
we proceed to construct the subdominant ultrametric, accordingly
with Equation (\ref{eq5}).
Firstly, we define the subdominant 
ultrametric distance matrix ${\bf D}^{<}$. This ultrametric 
matrix is obtained by defining the subdominant ultrametric 
distance $d(i,j)^{<}$ between countries $i$ and $j$, as the
maximum value of the distance $d(k,l)$ detected by moving,
in single steps, from country $i$ to country $j$ through 
the shortest path connecting $i$ and $j$ in the MST (Equation
\ref{eq5}). For instance, the ultrametric distance 
$d(SPA,POR)^{<}$ = 0.410 because both countries are placed 
side by side in the MST, and in this way, the ultrametric 
distance coincide with the metric distance, however, $d(SWI,SPA)^{<}$ =
0.490, which is the maximum metric distance between adjacent countries
in the path from SWI to SPA (see Figure 2 and Table 1). 
Proceeding in this way, we then order countries accordingly 
with their ultrametric distances to the others, placing the more
tightly connected countries in the center, and outward the
less connected. In order to obtain a clear picture of the
distances between countries, we have plotted in Figure (3)
the distance matrix given by Equation (\ref{eq3}), but 
countries ordered accordingly with their ultrametric distances. 

In the MST 
three groups of countries are clearly seen. It is interesting to 
note that these groups are built by geographical neighbor countries. 
EU countries group appears in first place with the smallest distances 
among them;
Asian countries followed and in third place Latin American 
countries have shown higher distances between their countries than 
the other two first groups. 
As expected, EU countries have shown the shortest distances in our 
sample (distances between 0.41 and 0.76) due to common 
relative real exchange movements\cite{AI00} inside 
the European Monetary System\footnote{In January 1, 1999, Spain, 
Portugal, Ireland, Finland, Italy and Norway, and other european 
countries not
in our sample, gave up their own currencies and adopted the Euro 
currency, with fixed nominal
exchange among them, and in January 2001 Greece joined the Euro 
too (United Kingdom, Denmark and Swedden refused to join the Euro). 
After Januay 1999, the Greek and Danish currencies joined to the 
new Exchange Rate Mechanism, where currencies are allowed to float 
within a range of $\pm 15 \%$ against the Euro. But as Greece joined 
the Euro in 2001, Denmark is the only participant in the mechanism 
in our country sample.},
although two different sub-groups of countries shows up; one in the north, with 
DEN as the most linked country and the other one in the south of 
Europe with short distances and intense links among SPA, POR and ITA. 
Finally, FIN is the least connected country in this group and U\_K and IRE 
seem not to belong to it.
Correlations coefficients 
in Figure (3) clearly support the closeness among EU 
countries exchange rate dynamics based, 
of course, in the common policy they have followed.

Asian countries form the next group order by distance. By far, 
THA and MAL are 
the most connected into the group (distances between 0,66 and 1) 
and are also quite connected 
with the EU countries and AUS and U\_K 
($1,05 < d < 1,3$). On the other side, 
KOR and especially INDI form a relatively isolated pair and have 
shown little and less intense
connections with any of the groups.  

\begin{figure}
\includegraphics[width=10cm,angle=270]{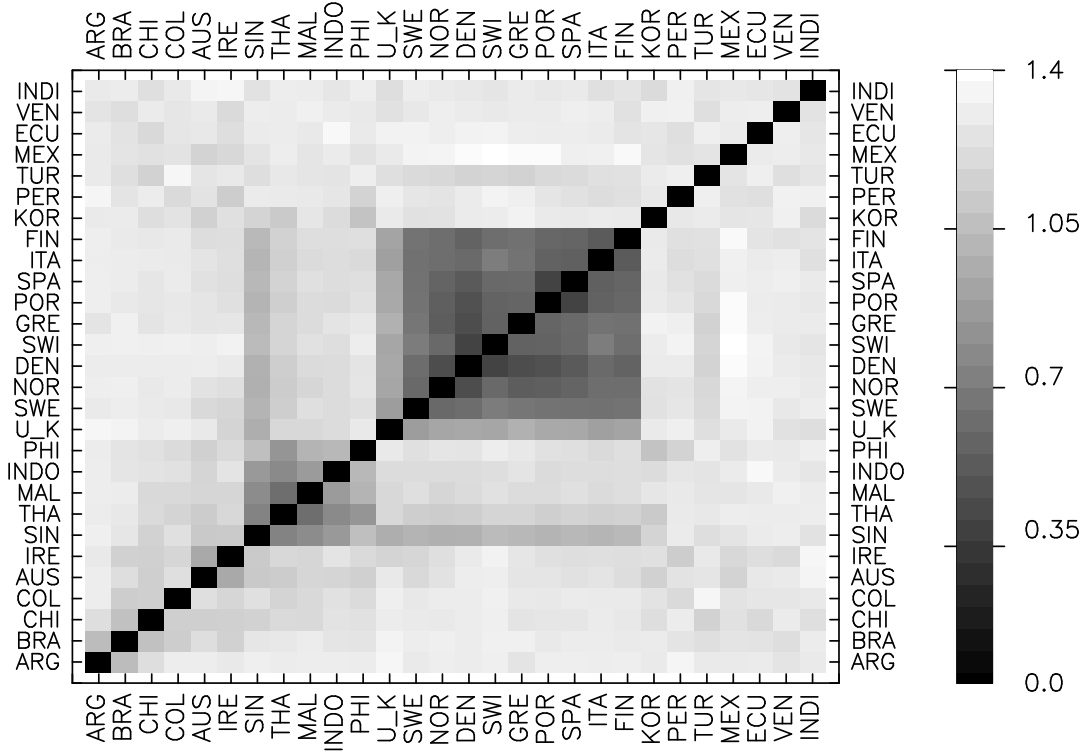}
\caption{
Gray scale distance plot. Distance measure is
calculated accordingly with Equation (\ref{eq3}). Countries
are ordered in the $x$ and $y$ axis accordingly with their
ultrametric distances (see texts), and they
are labeled accordingly with the symbols listed in the Appendix.
\label{f3}}
\end{figure}

Our third group is the Latin American one. Distances show 
high values (above 1,1) and very 
diffuse connections so, in fact, it is not a homogeneous group. 
Interesting enough is the important 
role played by BRA in South America as a centre of connections 
in this region. In this sense, 
BRA is the first link for ARG, CHI and 
COL showing the central role of their exchange 
rate economic policy in the South American continent. 
(In Figure (\ref{f3}) the correlation coefficients 
show the same central role of BRA). On the other hand, 
ARG, PER, ECU and VEN have shown relative isolated 
exchange rate dynamics in the analyzed period, with no apparent relevant 
links in the region. The same occurs to MEX but in this case the reason 
probably was their intense trade and financial relations with the United States.
In Figure (\ref{f3}) we can see no apparent group formation in the region, 
except light correlations in BRA.

In this regional hierarchy there are countries with connections more 
"diffuse". For instance, the U\_K shows small distances with 
the EU group (0,95) in first term but also with Asian countries and 
IRE and AUS. In the same direction, CHI shows short distance with 
BRA (1,17)  in first time but immediately are AUS, COL, IRE, 
ECU and MAL. More isolated is INDI with very 
diffuse connections and 
high distances (1,33), to KOR, 
U\_K, CHI, ITA and SIN.

\section{Conclusions}
We have introduced a new criterion to characterize the effects of 
currency crises  based solely on the correlations of real exchange 
rate returns time series. By using the information provided by
the correlations between synchronous movements in the real exchange
rates in different countries, we were able to construct a 
geometrical picture of the countries connections by means the MST.
Moreover, taxonomic information is also extracted from the time series,
by ordering countries
accordingly with its ultrametric distance (Figure \ref{f3}).

The hierarchical structure has shown three groups of 
countries which are clearly divided in a regional dimension. 
EU and Asian countries are relatively homogenous groups, meanwhile 
Latin American countries form a heterogeneous region where Brazil 
exchange rate dynamics is central. On the other side, we have shown a 
group of countries which do not belong to a specific group, 
such us Chile, India or United Kingdom.
From an economic point of view, information of our hierarchical tree 
could be useful in three relevant aspects.
First of all, we would expect that countries or group of countries with 
short distances among them were affected commonly by the same, 
or almost the same, economic and non economic factors, 
such as the EU group and in the central Asian group. When 
distances are larger among countries, exchange rate dynamics are 
affected by country specific factors.

In second place, information of our tree could be of interest for 
defining different methods of dating a currency crises depending of 
the range of countries to be used in the empirical analysis. 
So, this approach could improve results in dating a currency crises 
and also in defining the event of crises. In the same direction, 
this taxonomy can be used to define different regional or individual 
Early Warning Systems.

In third place, the taxonomy associated with the obtained hierarchical structure 
might be useful in the theoretical description of contagion and in the 
search of specific economic and no economic factors affecting different 
groups of countries. In addition, this hierarchy may be a useful tool
in the analysis of exchange rate crises contagion.

\section*{Acknowledgments}
We would like to thank, without implicating, International 
Economics Research Group in Oviedo University. 
G.O. thanks financial 
support from the Consejo Nacional de Investigaciones Cientificas 
y T\'ecnicas, Argentina. D. M. thanks financial support from the University of Oviedo.

\appendix
\section{Countries}
The 28 countries included in this work are as follows:
Argentine (ARG), Malaysia (MAL), Thailand (THA), Mexico (MEX), Korea (KOR), 
Indonesia (INDO), Brazil (BRA), Venezuela (VEN), Peru (PER), India (INDI), 
Ecuador (ECU), Turkey (TUR), Colombia (COL), Singapore (SIN), Philippines (PHI), 
United Kingdom (U\_K), Sweden (SWE), Italy (ITA), Ireland (IRE), Finland (FIN), 
Chile (CHI), Greece (GRE), Portugal (POR), Switzerland (SWI), Denmark (DEN), 
Spain (SPA), Norway (NOR), Australia (AUS)

%\section*{References}

\end{document}